\documentclass[aps,prd,a4paper,nosuperscriptaddress,nofootinbib,showpacs,twocolumn,showkeys,amsfonts,amssymb,amsmath]{revtex4-2}
\usepackage{amssymb,latexsym}

\usepackage{color}
\usepackage{amscd}
\usepackage{times}
\usepackage{epsfig}
\usepackage{psfrag}
\usepackage{graphicx}
\usepackage{amsthm,amsmath}
\usepackage{mathrsfs}
\usepackage[colorlinks=true, allcolors=blue]{hyperref}
\usepackage{leftindex}
\usepackage[dvipsnames]{xcolor}
\usepackage{physics}

\begin{document}
\title[]{Post-Newtonian Constraints on Semiclassical Gravity with Quantum Superpositions}

\author{Hollis Williams}
\affiliation{ Theoretical Sciences Visiting Program, Okinawa Institute of Science and 
Technology Graduate University, Onna 904-0495, Japan}

 \begin{abstract}
Semiclassical gravity, in which a classical spacetime is sourced by the quantum expectation value of the stress-energy tensor, is a standard framework for describing the gravitational interaction of quantum matter.  
In the nonrelativistic limit this approach leads to the Schrödinger-Newton equation, which is often assumed to be consistent at least in the weak-field regime.  In this work, we reexamine this assumption for spatial quantum superpositions of massive particles.  
We show that, when the quantum state is properly normalized, no modification of the Newtonian gravitational potential arises at leading order.  
However, at first post-Newtonian order the semiclassical coupling generically produces state-dependent contributions involving the mass density and the mass current of the superposition.  
These terms have a parametric scaling which is different from that of the corresponding relativistic corrections and which does not have Planck mass suppression.  Our results therefore impose a strong post-Newtonian consistency constraint on deterministic semiclassical gravity, indicating that sourcing the metric solely by expectation values is insufficient to recover a consistent relativistic weak-field expansion.

\end{abstract}

\maketitle

\section{Introduction}

Understanding how gravity couples to quantum matter remains a central open problem in fundamental physics \cite{rosenfeld, hu, kiefer}.  A standard approach is semiclassical gravity, in which the spacetime metric is treated classically whilst being sourced by the quantum expectation value of the stress-energy tensor

\begin{equation}
G_{\mu\nu} = 8\pi G \langle T_{\mu\nu} \rangle .
\end{equation}

\noindent
This framework arose naturally in early attempts to reconcile quantum theory with general relativity and provides a minimal setting in which quantum matter influences a classical gravitational field \cite{moller}.

In the non-relativistic limit, semiclassical gravity leads to the Schr\"odinger-Newton equation, where the Newtonian potential obeys the Poisson equation with a source proportional to the density $\rho(\textbf{x})= m|\psi(\textbf{x})|^2$ \cite{diosi, penrose, moroz}.  This model provides a simple setting for exploring how quantum spatial superpositions of massive particles influence the classical gravitational field without requiring a complete theory of quantum gravity.  Because the gravitational field depends on the quantum state, this equation introduces deterministic nonlinearity into quantum dynamics and has been widely studied in the context of macroscopic superpositions, gravitational self-interaction, and proposed mechanisms for wavefunction collapse \cite{giulini1, giulini2, yang}.   On the experimental side, there have been various proposals which search for indirect signatures of the Schr\"odinger-Newton equation in interferometry and optomechanical systems \cite{grossardt, gan, aguiar}.  However, despite extensive work, the internal consistency of this semiclassical framework beyond the Newtonian limit has not been systematically explored.

It is often implicitly assumed that if semiclassical gravity reproduces Newtonian physics for localized states, then relativistic corrections can be incorporated perturbatively into a standard post-Newtonian (PN) expansion.
However, post-Newtonian dynamics imposes strong structural constraints arising from Lorentz invariance, stress-energy conservation, and the effective field theory structure of general relativity \cite{goldberger, blanchet, poisson, porto, will}.  Whether a deterministic semiclassical coupling sourced by quantum expectation values rather than a classical stress-energy tensor can satisfy these constraints is therefore not obvious and has not been examined carefully.  There are of course alternative approaches to rendering semiclassical gravity consistent, such as conditional emission frameworks in which quantum systems only source a gravitational field when participating in specific interaction chains \cite{pipa}. These models seek to evade the problems which can arise from the conventional $\langle T_{\mu\nu} \rangle $ coupling, but do not directly address its internal consistency in the standard post-Newtonian regime.

In this work, we study the question of consistency for spatial quantum superpositions of massive particles.
We consider a simple but very general scenario where a massive particle is prepared in a spatial superposition of $N$ wavepackets, with $\Delta x_{ij}$ denoting the distance of separation between two wavepackets $i$ and $j$. 
We first show that, when the quantum state is properly normalized, no modification of the Newtonian gravitational potential arises at leading order: the Newtonian limit of semiclassical gravity is identical to that of general relativity.  We then demonstrate that the situation changes qualitatively as soon as one goes to first order in the PN expansion.  Since the metric is sourced by $\langle T_{\mu\nu} \rangle$, the semiclassical theory generically produces state-dependent PN contributions involving the mass density and the mass current of the quantum superposition.
These terms have a parametric scaling which is drastically different from the corresponding relativistic corrections and which importantly do not have Planck mass suppression.  As a result, the semiclassical theory fails to reproduce the universal state-independent post-Newtonian structure required by general relativity.

Our results therefore impose a strong consistency constraint on deterministic semiclassical gravity.  Although the Newtonian limit is preserved as expected, the theory does not appear to admit a relativistically consistent post-Newtonian extension when the gravitational field is sourced by quantum superpositions.  This identifies post-Newtonian dynamics as a powerful and previously unexplored probe of the interface between quantum mechanics and gravity.

\section{Newtonian Limit}

Before extending to the post-Newtonian regime, we will begin by considering the behavior of semiclassical gravity in the strict
Newtonian limit. In semiclassical gravity, the Newtonian potential satisfies
the Poisson equation
\begin{equation}
    \nabla^2 V(\mathbf r) = 4\pi G \, \langle \rho(\mathbf r) \rangle ,
\end{equation}
where the mass density operator $\rho$ is replaced by its quantum expectation value.  For a single non-relativistic particle of mass $m$ with wavefunction
$\psi(\mathbf x)$, one has

\begin{equation}
    \langle \rho(\mathbf x) \rangle = m |\psi(\mathbf x)|^2 .
\end{equation}

\noindent
Next, consider a general spatial superposition of localized wavepackets,
\begin{equation}
    \psi(\mathbf x) = \sum_i c_i \psi_i(\mathbf x),
\end{equation}
with each $\psi_i$ individually normalized and coefficients $c_i$ satisfying the usual condition
$\sum_i |c_i|^2 = 1$.  Normalization of the quantum state then guarantees conservation of the
total mass

\begin{equation}
    \int d^3x \, \langle \rho(\mathbf x) \rangle
    = m \sum_i |c_i|^2 = m .
\end{equation}

At distances $r$ which are large compared to the spatial extent of the wavefunction, the
Newtonian potential is determined solely by this monopole moment. Expanding
the Green function for $r \gg |\mathbf x|$ via a Taylor expansion yields

\begin{equation}
    V(r) = -G \int d^3x \, \frac{\langle \rho(\mathbf x) \rangle}{|\mathbf r-\mathbf x|}
    = -\frac{Gm}{r} + \mathcal{O}(r^{-2}).
\end{equation}

\noindent
This is independent of the coherence, separation, or number of wavepackets in the
superposition.  Crucially, interference terms in $|\psi(\mathbf x)|^2$ modify the spatial distribution
of the mass density, but do not alter its monopole moment.  They therefore cannot produce an observable correction to the Newtonian potential at leading
order. Any apparent renormalization of Newton's constant would violate conservation of mass and arises only if the quantum state is normalized incorrectly.  The subleading $\mathcal{O}(r^{-2})$ terms in this expansion correspond to the usual
finite-size multipole corrections arising from the spatial extent of the mass
distribution.  These terms would be present for any extended classical source and do not
constitute an observable modification of Newton’s constant $G$.

This result shows that semiclassical gravity with properly normalized quantum
states reproduces the classical Newtonian limit exactly. It therefore follows that any genuine
deviation from classical gravity within semiclassical models must arise only when we extend beyond the Newtonian approximation. In the following section, we demonstrate that such
deviations instead appear naturally at post-Newtonian order, where the gravitational
field becomes sensitive to state-dependent mass currents and relativistic
corrections that cannot be absorbed into a redefinition of Newton's constant.

\section{Post-newtonian inconsistency}

In the weak-field regime with slowly moving particles, both general relativity and alternative
theories of gravity produce post-Newtonian corrections to the Newtonian potential with the generic
form

\begin{equation}
    V_{1\mathrm{PN}} \;=\; \mathcal{C}\frac{G m_s m_t}{r}\frac{v^2}{c^2},
\end{equation}
where $\mathcal{C}$ is a dimensionless coefficient, $m_s$ is the source mass, $m_t$ is the test mass, and $v$ is the characteristic relative velocity.  In general relativity, this coefficient is of order unity and does not depend on the structure or the quantum state of the source.

However, in a semiclassical Diósi-type violation of quantum mechanics, the 
metric is sourced by the expectation value $\langle T_{\mu\nu} \rangle$, so that the 
$1$PN terms acquire contributions which depend explicitly on the quantum state.  These include the mass density $\rho(\mathbf x)$ and the mass current $j(\mathbf x)$ of the
superposed source mass.  Although the absolute $1$PN corrections in both theories contain the same prefactor 
$G m_s m_t/r$, this can be removed by taking the ratio of the potentials $V^{\mathrm{(SC)}}_{1\mathrm{PN}}/V^{\mathrm{(GR)}}_{1\mathrm{PN}}$, isolating the physical difference between general relativity and the semiclassical theory.  This ratio isolates a dimensionless coefficient characterizing the post-Newtonian structure of the theory, in the same form constrained by precision tests of relativistic gravity.

The simplest starting point is to consider the representative $1$PN box diagram containing one insertion of the
mass current $T^{0i}$ and one insertion of the mass density $T^{00}$ on the 
quantum source.  
In general relativity this diagram, combined with the other $1$PN contributions, gives the usual Einstein-Infeld-Hoffmann potential, whose scaling is
\begin{equation}
    V_{1\mathrm{PN}}^{\mathrm{(GR)}} 
    \;\sim\; \frac{G^2 m_s^2 m_t}{c^2 r^2}.
\end{equation}

\noindent
In the semiclassical Diósi-type theory, the same diagram is obtained by 
replacing the classical stress-energy insertions by their expectation values 
$\langle T^{\mu\nu} \rangle$.  
This turns the box diagram into a convolution of the mass density 
$\rho(\mathbf x)$ and mass current $j(\mathbf x)$ of the quantum state:
\begin{equation}
    V_{1\mathrm{PN}}^{\mathrm{(SC)}}
    \;\sim\;  G^2 m_t
    \int d^3x\, d^3x'\,
    \frac{j^i(\mathbf x)\,\partial_{x'_i}\rho(\mathbf x')}
         {|\mathbf x - \mathbf r|\,|\mathbf x' - \mathbf r|}.
\end{equation}

For a Gaussian superposition of width $\sigma$, the density scales as $ 1/\sigma^3$, the gradient scales as $1/\sigma^4$, and the 
mass current scales as $\hbar/(m_s \sigma^4)$.  
The spatial integrals over the normalized Gaussians give quantities of order unity, so the overall semiclassical contribution scales as
\begin{equation}
    V_{1\mathrm{PN}}^{\mathrm{(SC)}} 
    \;\sim\; 
    \frac{G^2 m_t}{r^2}\,\frac{\hbar}{m_s \sigma^2}.
\end{equation}

\noindent
Although we have assumed a Gaussian wavepacket to make the scaling explicit,
the argument does not rely on this specific choice.
For any spatially extended quantum state characterized by a single length
scale $\sigma$, dimensional analysis implies a typical momentum scale
$p \sim \hbar/\sigma$ and hence a mass current
$j \sim \rho\, p/m_s \sim \hbar/(m_s \sigma^2)$.
The appearance of $\hbar$ and inverse powers of $\sigma$ in the 1PN sector
is therefore generic for spatial quantum superpositions and does not depend
on the functional shape of the wavepacket.

Taking the ratio of the two $1$PN contributions yields the dimensionless quantity
\begin{equation}
\label{eq:ratio}
    \frac{V_{1\mathrm{PN}}^{\mathrm{(SC)}}}
         {V_{1\mathrm{PN}}^{\mathrm{(GR)}}}
    \;\sim\;
    \frac{\hbar c^2}{m_s^3 \sigma^2}.
\end{equation}
Importantly, this coefficient does not have any Planck mass suppression.  For 
representative laboratory parameters (e.g.\ $m_s \sim 10^{-12}$\,kg, 
$\sigma \sim 50\,\mu\mathrm{m}$) one finds
\begin{equation}
    \frac{V_{1\mathrm{PN}}^{\mathrm{(SC)}}}
         {V_{1\mathrm{PN}}^{\mathrm{(GR)}}}
    \;\sim\; 10^{27}.
\end{equation}

\noindent
This enormous factor shows that the semiclassical $1$PN contribution is 
many orders of magnitude larger than the GR $1$PN term for a typical superposition of masses.  This estimate was obtained from a representative $1$PN box diagram involving one
mass current and one mass density insertion.  One may still ask if the full sum of all the 1PN diagrams in the semiclassical theory could fortuitously cancel out this contribution and rescue consistency.  We will now argue that such a cancellation is not possible on general grounds.

In general relativity, all $1$PN corrections arise from classical point-particle
stress-energy tensors and have polynomial scaling in the masses and velocities of the sources.
In particular, the Einstein-Infeld-Hoffmann potential only  depends on $m_s$, $m_t$,
and $v^2/c^2$, and contains no intrinsic length or action scales.  By contrast, in the semiclassical theory the metric is sourced by expectation values
$\langle T^{\mu\nu} \rangle$ evaluated on a quantum state.  The $1$PN sector therefore
contains contributions involving the mass current
$j^i(\mathbf x) = \langle \hat T^{0i}(\mathbf x) \rangle$.

For a spatial superposition of width $\sigma$, this current scales as
$j \sim \hbar/(m_s \sigma^2)$, introducing an explicit factor of $\hbar$ into the
post-Newtonian expansion.
At 1PN order, the allowed operator structures are constrained by
Lorentz invariance, stress-energy conservation, and power counting.  Within the standard post-Newtonian EFT power counting and stress–energy conservation, no other allowed 1PN contribution consistent with these constraints carries the same dependence on
$\hbar$, $m_s$, and the superposition scale $\sigma$.
In particular, diagrams involving only mass-density insertions depend solely
on classical monopole and multipole moments and are insensitive to quantum
coherence.  Consequently, there is no possible counterterm at $1$PN order which can cancel out the contribution identified above.

We therefore conclude that the ratio given in equation (11) is a robust prediction of naive semiclassical coupling, and not merely a factor associated with a particular Feynman diagram which is lost when one calculates the full sum.  On the experimental side, post-Newtonian deviations from general relativity are tightly constrained by
Solar System tests and binary pulsar observations \cite{damour, kramer, clifford}.  These tests establish that
the 1PN sector of gravity is described by universal, state-independent
coefficients of order unity.
These constraints do not rely on the mass scale of the source, but rather
on the structural form of the post-Newtonian expansion itself.  By contrast, the enhancement factor found in equation (12) arises whenever a
source is placed in a spatial quantum superposition of width $\sigma$,
even for extremely small laboratory scale masses \cite{bose1, vedral, schut2, schut1}.  The appearance of a state-dependent 1PN coefficient exceeding unity by
many orders of magnitude therefore signals a failure of the semiclassical
theory to reproduce the universal post-Newtonian structure.

This conflict reflects an internal inconsistency of deterministic
semiclassical gravity and is therefore in direct conflict with existing bounds.  The inconsistency follows directly from the power-counting structure of the post-Newtonian expansion once the metric is sourced by quantum expectation values.  It is important to emphasize that the effect identified here relies essentially on quantum coherence: for a single localized wavepacket, or for an incoherent statistical mixture, the expectation value of the mass current reduces to its classical form and no anomalous post-Newtonian contribution arises.

\section{Conclusion}

We have examined the coupling between classical gravity and spatial quantum superpositions within a deterministic semiclassical framework of Diósi–Schrödinger–Newton type.  
We showed that when the quantum state is properly normalized, there is no observable modification of Newton’s constant at leading order.  This demonstrates that the Newtonian limit of semiclassical gravity is more subtle than often assumed, and that apparent modifications of the gravitational coupling can be artifacts of an inconsistent treatment of quantum superpositions.

However, the situation changes qualitatively at post-Newtonian order.  At first order, the semiclassical sourcing of the metric by the expectation value $\langle T_{\mu\nu} \rangle$ generically introduces state-dependent contributions involving the mass density and the mass current of the quantum superposition.  We then showed that these contributions scale differently from the corresponding relativistic terms and that they can exceed the standard $1$PN corrections by many orders of magnitude for experimentally relevant parameters.  Crucially, this enhancement is not suppressed by the Planck mass, indicating that it reflects a structural inconsistency rather than a small quantum gravity correction.  Indeed, Planck mass suppression is likely the only mechanism which could bring this contribution down to a small size.

Our results therefore establish a strong consistency-based obstruction on naive deterministic semiclassical gravity.  Although such models may reproduce the Newtonian limit, they fail to admit a consistent post-Newtonian extension compatible with the EFT structure of general relativity.  Any viable semiclassical theory of gravity must therefore incorporate additional ingredients to restore consistency: these could be stochasticity, backreaction, or genuine quantum gravitational degrees of freedom.  Unlike earlier objections based on superluminal signaling or measurement-induced effects, the inconsistency identified here arises entirely at the level of classical relativistic dynamics
and does not rely on entanglement, observers, collapse postulates, or particular interpretations of quantum mechanics.

\section*{Acknowledgments}

\noindent
The author thanks Anupam Mazumdar for discussions and earlier work on the Newtonian formulation of this problem which motivated the post-Newtonian extension presented here.  This research was conducted whilst the author was visiting the Okinawa Institute of Science and 
Technology (OIST) through the Theoretical Sciences Visiting Program (TSVP).


\end{document}